# Modeling and simulation of polycrystalline ZnO thin-film transistors

Faruque M. Hossain[a), b)], J. Nishii[c)], S. Takagi[b)], A. Ohtomo, T. Fukumura
*Institute for Materials Research, Tohoku University, Aoba-ku, Sendai 980-8577, Japan.*
H. Fujioka
*Department of Applied Chemistry, The University of Tokyo, Bunkyo-ku, Tokyo 113-8656, Japan.*
H. Ohno
*Research Institute of Electrical Communication, Tohoku University, Sendai 980-8577, Japan.*
H. Koinuma
*Materials and Structures Laboratory, Tokyo Institute of Technology, Yokohama 226-8503, Japan.*
M. Kawasaki
*Institute for Materials Research, Tohoku University, Aoba-ku, Sendai 980-8577, Japan.*

Thin film transistors (TFTs) made of transparent channel semiconductors such as ZnO are of great technological importance, because their insensitivity to visible light makes device structures simple. In fact, several demonstrations are made on ZnO TFT achieving reasonably good field effect mobilities of 1-10 $cm^2$/Vs, but reveal insufficient device performances probably due to the presence of dense grain boundaries. We have modeled grain boundaries in ZnO thin film transistors (TFTs) and performed device simulation using a two-dimensional device simulator for understanding the grain boundary effects on the device performance. Actual polycrystalline ZnO TFT modeling is commenced with considering a single grain boundary in the middle of the TFT channel formulating with a Gaussian defect distribution localized in the grain boundary. A double Shottky barrier is formed in the grain boundary and its barrier height are analyzed as functions of defect density and gate bias. The simulation is extended to the TFTs with many grain boundaries to quantitatively analyze the potential profiles developed along the channel. One of the big contrasts of polycrystalline ZnO TFT compared with a polycrystalline Si TFT is that much smaller nanoscaled grain size induces heavy overlap of double Shottky barriers. Through the simulation, we can estimate the total trap state density localized in the grain boundaries for a polycrystalline ZnO by knowing apparent mobility and grain size in the device.

---
[a)]Electronic mail: fhossain@imr.tohoku.ac.jp
[b)]On leave from Department of Innovative and Engineered
Materials, Tokyo Institute of Technology, Yokohama
226-8502, Japan.
[c)]Also at Research Institute of Electrical Communication,
Tohoku University, Sendai 980-8577, Japan.

## I. INTRODUCTION

ZnO has become an attractive wide band gap semiconductor since ultraviolet laser action was demonstrated at room temperature.[1-3] Due to its transparent nature, it is of great importance to realize a transparent thin film transistor (TFT), which would be useful for driving an active-matrix liquid-crystal displays (AMLCDs). By substituting the TFTs made of amorphous Si (a-Si) or poly crystalline Si (poly-Si) currently used in AMLCDs by transparent ZnO TFTs, one can improve the opening of the pixels to reduce power consumption and avoid complicated device processing. There have been several experimental reports on the ZnO TFTs,[4-8] among which we have demonstrated reasonably high field effect mobility ($\mu_{FE}$) of 7 cm$^2$/Vs (Ref. 8) in ZnO TFT having device structure, dimensions, process temperature, and materials set compatible to those of a-Si TFTs. In terms of $\mu_{FE}$, which is higher than that of a-Si TFT ($\sqcup$ 1 cm$^2$/Vs), the performance of ZnO TFT is good enough for practical application. However, we need to understand the material chemistry and device physics to solve a few drawbacks such as wide sub-threshold characteristics and imperfect channel saturation.

Transparent and conducting polycrystalline ZnO thin films with the (0001) oriented texture can be deposited on glass or flexible plastic substrate using pulse laser deposition (PLD) at lower temperatures ($\sqcup$ 150$^o$C)[8] than that ($\sqcup$ 300$^o$C) employed for a-Si deposition in commercial AMLCDs. These films are composed of nano-crystals with grain size ranging from 50 to 100 nm and, consequently, contain high density of grain boundaries (GBs). This dense GB formation has an advantage of having semi-insulating film due to overlapped depletion regions, as we will explore in this paper. The GBs in ZnO, in general, contain wide distribution of deep level traps,[9,10] which are considered to be the main constraints for such improper characteristic performances of ZnO TFTs. We have realized that the device simulation will be the initial basic tool for analyzing the activities of trap states in GBs and its effect on TFT performance. Consequently, we will be able to understand the key factors to improve the performance of ZnO based TFTs. No one has reported the effect of GBs on the performance of ZnO-based TFTs and even there is no concrete evidence on the defect distribution along the band gap inside the GBs for TFT grade ZnO thin film. Here, we report on the two-dimensional (2D) device simulation of ZnO TFT using Atlas device simulator.[11]

Enormous experimental research and theoretical calculations have been performed on poly-Si based TFTs, where some works on GB modeling and simulation[12-19] and its activity in TFTs performance are reported. Three basic properties *i.e.* wide band gap, small grain size, and possible existence of deep level trap states at GBs make these low temperature deposited ZnO films qualitatively different from that of conventional poly-Si thin films. The significant differences between the ZnO-based TFT and the poly-Si TFT modeling can be pointed out as: (i) in nanocrystalline ZnO thin film, the developed depletion regions around closely spaced GBs overlap to make potential profiles in the film different from that of microcrystalline poly-Si thin film, (ii) due to wide band gap of ZnO and possible existence of deep level traps in GBs, the GB potential barrier height modulation with gate bias for ZnO TFTs will also be different from that of poly-Si TFTs. In development history of poly-Si TFT, inclusive efforts have been paid to avoid GBs in the channel, since the low density of GBs gives large scatter of device performances depending on the number and location of GBs in the channel. The GBs are so uniformly distributed in ZnO-TFT that one may be able to overcome the problem with rather having GBs, if the characteristics of GBs can be well controlled through the understanding of their physical properties and suitable chemical modification of them.

In our model, the defects in the crystallite and the ZnO-gate insulator interface are assumed to be segregated to the GBs *i.e.* all the defects are localized in GBs. This assumption makes our model for GBs in ZnO thin film equivalent to GBs in ZnO varistors. Spectroscopic analyses on ZnO varistors[9,10,20,21] and ZnO bulk crystals[22] have detected the trap levels located as a deep states with ambiguous discrete energy levels even though the exact level detection is unresolved and depends on the accuracy of spectroscopic measurements. Therefore, it is reasonable here to assume a wide range of continuous defects distribution in the GB with a peak density at the mid gap. Here, our discussion will focus on the developed potential profile in the active channel layer and its effect on the properties of ZnO-TFTs, which will reflect the real microscopic view of ZnO thin film for TFT actions. We are also able to adequately estimate the unknown trap state densities inside the GBs comparing with the experimental values of field effect mobility $\mu_{FE}$ and grain size $l_g$ using this model.

## II. SIMULATION METHODOLOGY AND MATERIAL PARAMETERS

In the device simulator[11], the semiconductor parameters are defaulted for silicon. We need to specify the required parameters of ZnO as an active layer material in a TFT structure for ZnO-based TFT simulation. An experimentally optimized dimensions[8] [channel length $L = 5$ μm, channel width $W = 25$ μm, active channel layer thickness $t_{ZnO} = 100$ nm, and gate insulator thickness $t_{SiN} = 350$ nm] of a TFT structure for two-dimensional (2D) device simulation is considered as schematically given in Fig. 1 to get the best TFT performance as required for practical application. Except for ZnO channel, all material sets and dimensions are identical to those in the a-Si TFT, used in commercial AMLCDs. All necessary material constants of ZnO have been collected from different sources[23-29] except for two constants ($E_1$ and $E_2$) related to the Gaussian distribution of the defects in the GB as enumerated in Table I. The values of $E_1$ and $E_2$ are assumed material constants, to be used all over this modeling. All energy levels defined in Table I can be graphically represented by an energy level diagram including a GB energy band bending profile as illustrated in Fig. 2 and are referenced to the valence band edge $E_v = 0$. Hence, the GB energy barrier can be defined as, $qV_b = E_{cgb} - E_c = E_{vgb} - E_v$ and the conduction activation energy away from the GB (in flat band region) is represented as, $qV_n = E_c - E_f$ as clarified in Fig. 2, where $E_{cgb}$, $E_c$, and $E_{vgb}$ are maximum conduction band energy at the GB, minimum conduction band edge energy in crystallite, maximum valence band energy at the GB, respectively. The gate insulator Si$_3$N$_4$ is a default material in the simulator with given permittivity $\varepsilon_I = 7.55$.

TABLE I: Collected and assumed material parameters (constants) of ZnO for 2D device simulation.

| Material constants for ZnO | Values | References |
|---|---|---|
| Bandgap, $E_g$ (300K) ($E_c - E_v$) | 3.4 eV | 23 |
| Effective mass of electron in the conduction band, $m_e^*$ | 0.318 $m_o$ | 24 |
| Effective mass of hole in the valence band, $m_h^*$ | 0.5 $m_o$ | 25 |
| Dielectric constant, $\varepsilon_s$ | 8.12 | 24 |
| Hall mobility, $\mu_H$ | 150 cm$^2$/Vs [§] | 26, 27 |
| Electron affinity, $\zeta$ ($E_{vac} - E_c$) | 4.29 eV | 28 |
| Work function, $\phi_s$ ($E_{vac} - E_f$) | 4.45 eV | 28 |
| Donor level, $E_c - E_d$ | 30 meV | 29 |
| Energy level of peak trap state density in GB, $E_1$ [($E_{cgb} - E_{vgb}$)/2] | 1.7 eV | assumed |
| Characteristic decay energy of Gaussian distribution, $E_2$ | 0.25 eV | assumed |

[§]Reference 26, 27 reported the Hall mobilities for single crystal ZnO are 100 and 200 cm$^2$/Vs, respectively, and we use an averaged value of 150 in our modeling.

Other material parameters for a polycrystalline ZnO as specified in Table II i.e. donor density $N_d$ (cm$^{-3}$), grain size $l_g$ (μm), and total areal trap state density in GB $N_t$ (cm$^{-2}$) are considered as variables to analyze the device characteristics. The variation of $N_d$ is analogous to the generation of carriers by doping, which is subsequently equivalent to the accumulation or depletion of carriers through the application of gate voltage in the ZnO-based n-channel TFTs. The variation of $l_g$ and $N_t$ correspond to the deposition condition of ZnO films and its imperfection, respectively.

The polycrystalline ZnO thin film is defined by introducing equally spaced grain boundaries (GBs) parallel to the ZnO film thickness and perpendicular to the direction of carrier propagation from source to drain (see Fig. 1). The GB is modeled as a thin layer (i.e. a few atomic layers) having defect states with Gaussian distribution as[11,30];

$$N_{ga}(E) = N_{ta} \exp\left\{-\left[(E_{1a} - E)/E_{2a}\right]^2\right\} \quad (1)$$

$$N_{gd}(E) = N_{td} \exp\left\{-\left[(E - E_{1d})/E_{2d}\right]^2\right\} \quad (2)$$

$$N_g(E) = N_{ga}(E) + N_{gd}(E) \quad (3)$$

where subscripts g, a, and d stand for Gaussian distribution, acceptor-like states, and donor-like states, respectively. $N_g(E)$ and E are the density of the defect states and its corresponding trap energy inside the band gap, respectively. $N_t$, $E_1$, and $E_2$ stand for the total density of the trap states, its peak energy, and its characteristic decay energy, respectively. While, the slab between two GBs (crystallite) and ZnO-Si$_3$N$_4$ interface are assumed to be completely defect-free. This means, we assume all the defect states are localized in GBs. For a minimum mesh[31] size of 10 nm (Ref.12), an areal interface trap density, for example, 10$^{11}$ cm$^{-2}$ along the GB is equivalent to the volume density of 10$^{17}$ cm$^{-3}$ in the GB region. We assume an idealized situation that the distribution of both acceptor-like and donor-like defects are the same within the energy gap i.e. $N_{ta} = N_{td} = N_t$, $E_{1a} = E_{1d} = E_1$, and $E_{2a} = E_{2d} = E_2$. In the simulator, it is assigned that a donor-like trap is positively charged and therefore can only capture an electron i.e. donor-like traps are positive when unoccupied of an electron, but are neutral when occupied. On the other hand, an acceptor-like trap is negatively charged and therefore can only emit an electron i.e. acceptor-like traps are negative when occupied, but are neutral when unoccupied. The capture and emission processes are handled by the simulator using Shockley-Read-Hall (SRH) recombination model.[11] Trapping model described above is elaborately analyzed by Simmons and Taylor[32] and reported in J. Werner et al.[33] for an n-type semiconductor. Therefore, the negative and positive charges in GB created by trap states are determined by the following probability equations;

$$n_a = \int_{E_v}^{E_c} N_{ga}(E) f(E) dE \quad (4)$$

$$n_d = \int_{E_v}^{E_c} N_{gd}(E) [1 - f(E)] dE \quad (5)$$

TABLE II: Used material parameters (variables) of ZnO for 2D device simulation.

| Material variables for ZnO | Symbol (unit) | Values |
|---|---|---|
| Donor density | $N_d$ (cm$^{-3}$) | $5.0 \times 10^{14} - 1.0 \times 10^{17}$ |
| Grain size | $l_g$ (μm) | 0.2 – 1.0 |
| Total areal trap state density in GB (both donor-like and acceptor-like) | $N_t$ (cm$^{-2}$) | $5.0 \times 10^{11} - 1.0 \times 10^{13}$ |

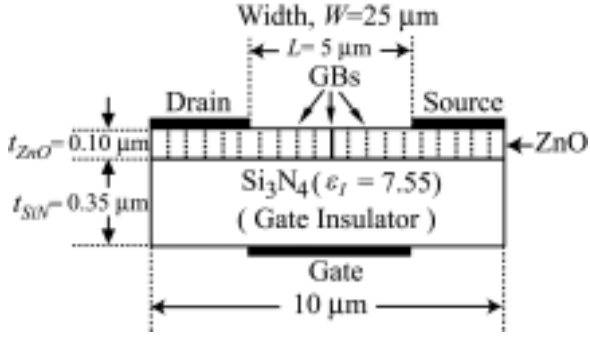

FIG. 1. Schematic cross-sectional structure of a ZnO TFT for 2D device simulation.

$$f(E) = \frac{1}{1+\exp[(E-E_f)/kT]} \quad (6)$$

where $f(E)$ is the occupation probability of defect states, $E_f$ is the equilibrium Fermi level, $k$ is the Boltzmann's constant, $T$ is the lattice temperature, $n_a$ and $n_d$ are the occupied acceptor-like trap state density and unoccupied donor-like trap state density in the GB, respectively. The value of $T$, considered in this model, is 300K i.e. all calculations are performed at room temperature. Hence, the Poisson's equation to determine the potential profiles along the ZnO channel in the defect-free crystallite regions and in the GB regions are

$$\frac{\partial^2 V}{\partial x^2} = -\frac{q(n-p-N_d^+)}{\varepsilon_s} \quad (7)$$

and

$$\frac{\partial^2 V}{\partial x^2} = -\frac{q(n-p-N_d^+)}{\varepsilon_s} + \frac{qn_t}{\varepsilon_s} \quad (8)$$

respectively, where $n$, $p$, $N_d^+$, and $n_t = n_a - n_d$ are the electron density, the hole density, the ionized shallow donor density, and the net negative charge in the GB, respectively. This net negative charge $n_t$ depletes neighboring electrons from GB and causes to expose the ionized shallow donors

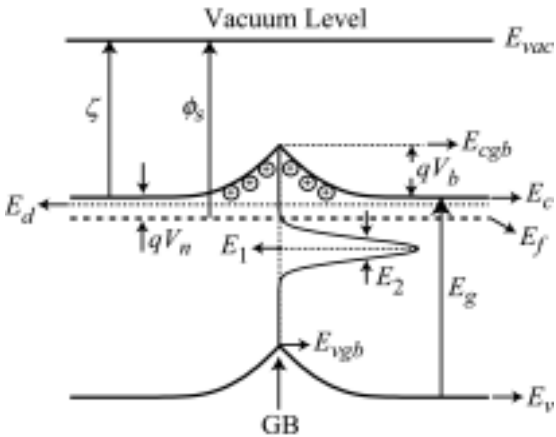

FIG. 2. Energy band diagram for all defined energy levels correspond to Table I in presence of a grain boundary. All levels are referenced to minimum energy level $E_v = 0$ and maximum energy level $E_{vac}$.

$N_d^+$, which in turn results in a band bending around the GB.

Consequently, a double Shottky barrier is formed at electrostatic and thermal equilibrium. This double Shottky barrier is the main obstacle for the carriers to transport through the polycrystalline ZnO channel. Two different transport mechanisms are utilized for the GB and for the defect free crystallite regions without any approximation in the device simulator. The thermionic emission process and drift-diffusion process are used for carrier transport over GB potential barrier and for defect free crystallite, respectively.

All the physical events described above for TFT actions can be numerically analyzed by the device simulator using the following required equations;

$$n = N_c \exp[-(E_c - E_f)/kT] \quad (9)$$

$$p = N_v \exp[-(E_f - E_v)/kT] \quad (10)$$

$$N_d^+ = N_d (1/\{1+\exp[(E_f - E_d)/kT]\}) \quad (11)$$

$$\vec{J} = qn\mu\vec{E} + qD\nabla n \quad (12)$$

$$\vec{J}_{gb} = AT^2 \frac{n}{N_c}\exp\left(-\frac{qV_b}{kT}\right) \quad (13)$$

where $N_c$ and $N_v$ are the effective density of states (DOS) for electron in conduction band and the effective DOS for holes in the valence band, $N_d$ and $E_d$ are the shallow donor density and its corresponding energy level, $\vec{J}$ and $\vec{J}_{gb}$ are the electron current densities in the crystallite and in the GB. $\mu$, $\vec{E}$, $\nabla n$, $D$, and $A$ are the electron mobility, the effective electric field, the electron concentration gradient, the diffusion coefficient for electron, and the electron Richardson constant, respectively. Equations (9) and (10) are the carrier density equations for electrons and holes, respectively. Equation (11) is the occupation probability equation for shallow donors. Equations (12) and (13) are the carrier current density equation to describe the drift-diffusion process in the crystallite and the thermionic emission process in the GB, respectively.

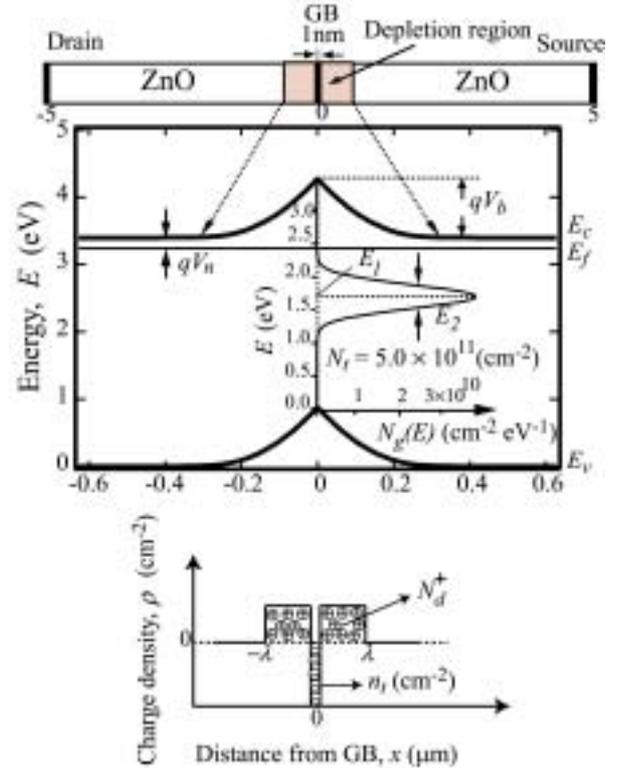

FIG. 3. Energy band bending model of a single grain boundary (SGB) with Gaussian type defect distribution inside the band gap of ZnO, assuming total areal trap state density localized at GB $N_t = 5.0 \times 10^{11}$ cm$^{-2}$, its peak trap density energy level $E_1 = 1.7$ eV, and its characteristic decay energy $E_2 = 0.25$ eV.

Subsequent equations to describe the device characteristics will be based on the above basic equations, which can also be efficiently operated by the device simulator without any approximation.

## III. RESULTS AND DISCUSSION
### A. Single grain boundary (SGB) modeling

A SGB structure of ZnO, set up with materials constants taken from Table I, is modeled in Fig. 3 and can be treated as an $n$-type bi-crystal. An 1nm-thick layer is considered as a GB introducing Gaussian defect distribution distributed potential $V(x)$ and corresponding electron concentration $n(x)$ are described as a function of distance $(x)$ from GB to space charge regions as;

$$V(x) = V_b \exp(-\frac{|x|}{\lambda}) \quad (14)$$

$$n(x) = n \exp\left\{-\frac{qV(x)}{KT}\right\} \quad (15)$$

where $\lambda$ is the characteristic decay length of potential for gradual depletion and $n$ is the free carrier concentration at thermal equilibrium in the crystallite regions. The values of $\lambda$ ranges $L_d < \lambda < W_d$, where $L_d$ is the Debye length and $W_d$ is the depletion width, which are usually expressed by the equations

$$L_d = (\varepsilon_s kT / q^2 n)^{1/2} \text{ and } W_d = (\varepsilon_s V_b / qn)^{1/2} \quad (16)$$

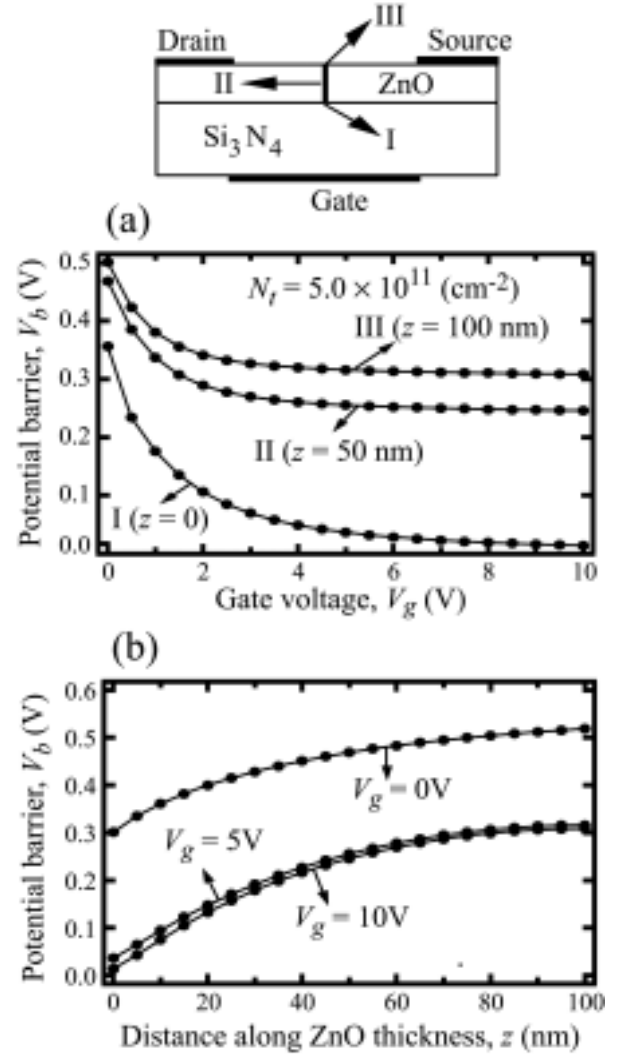

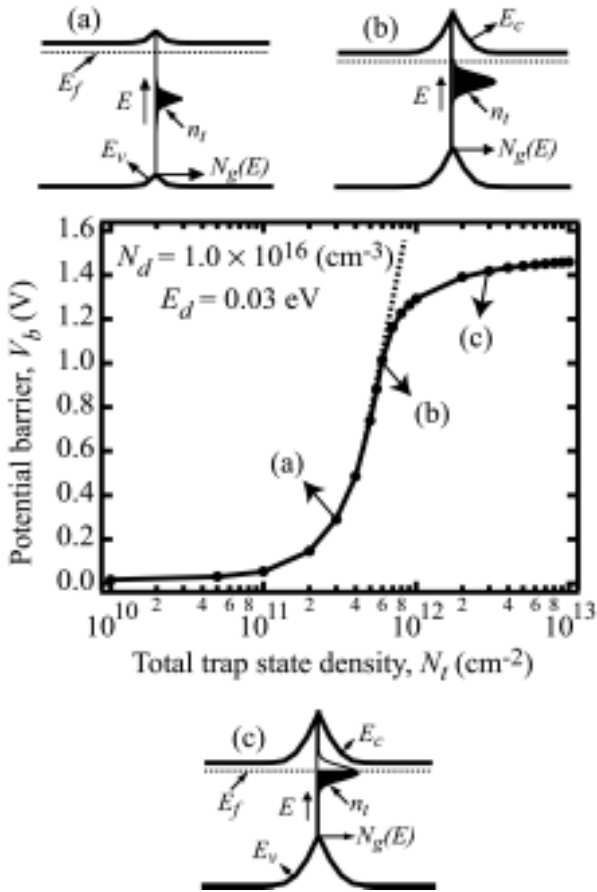

FIG. 4. Change in potential barrier height $V_b$ of a SGB with the total areal trap state density $N_t$ for a fixed donor density $N_d = 1.0 \times 10^{11}$ cm$^{-3}$ and donor level $E_d = 0.03$ eV. The schematics of three band bending profiles (a), (b), and (c) are drawn corresponding to three different points of the $V_b - N_t$ plot. The dotted line is given by equation (17), where all the traps are occupied as shown in (a).

FIG. 5. Change in potential barrier $V_b$ in a single grain boundary (SGB) (a) with gate voltage $V_g$ at three locations along the ZnO thickness (I: at the ZnO-Si$_3$N$_4$ interface, II: at the middle of ZnO film, and III: at the surface of ZnO film) (b) along the ZnO thickness for three different gate bias conditions ($V_g = 0, 5,$ and 10 V).

using equations (1) to (3). To simplify the model, we assume $E_1 = 1.7$ eV, $E_2 = 0.25$ eV, and the same distribution for both acceptor-like and donor-like defects. Developed respectively, where $\varepsilon_s$ is the permittivity of ZnO. The net areal negative charge $n_t$ (cm$^{-2}$) at GB can be estimated using abrupt depletion approximation and charge neutrality condition, $n_t = 2\lambda N_d^+$ (Ref. 34). For a single GB case, grain size $l_g$ is large compared to the abrupt depletion length $\lambda$ and all the trap states are occupied ($n_t = N_t$), because the Fermi level is above the all trap levels (Fig. 3). In this case, the electrons are depleted partially and developed potential barrier $V_b$ follows a square relationship with $N_t$ (cm$^{-2}$) as;

$$V_b = qN_t^2/8\varepsilon_s n = qn_t^2/8\varepsilon_s n \qquad (17)$$

The variation of $V_b$ with $N_t$ is a crucial factor to limit the carrier transportation, causing a remarkable change in device performance. Therefore, it is important to explore the maximum limit of $V_b$ with $N_t$ values at SGB. Figure 4 points up the variation of $V_b$ with $N_t$ for a fixed donor density $N_d = 1.0 \times 10^{16}$ (cm$^{-3}$) and donor level $E_d = 0.03$ (eV). Below a certain value of $N_t$, $V_b$ follows the equation (17) given by dotted line, because all the trap states are occupied and the established net negative charge at GB ($n_t$) remains equal to $N_t$ as illustrated by a schematic band diagram (a). As $N_t$ increases, the calculated curve (solid line) starts to deviate from the dotted line at a critical value of $N_t$ [(b)]. At much higher $N_t$ values, the trap density levels traverse up the Fermi level resulting in creating unoccupied trap states above the Fermi level. The term $N_t$ is then replaced by the term $n_t$ as was described in equation (17). Hence, the $V_b$ solely depends on the net negative charge $n_t$ in the GB. As the Fermi level impinges deeply into the trap states as indicated by a schematic band diagram (c), the $n_t$ values start to saturate and correspondingly so does $V_b$. This is due to a huge compromise between equations (4) to (6) and (11). Finally, $V_b$ reaches to its maximum value self-consistently under thermal equilibrium condition. It is worth to mention that the values of $n_a$, $n_d$, and $N_d^+$ depend on the deposition condition of the ZnO films and the quality of the substrate on which the ZnO films are grown.

To investigate the effect of GBs on TFT properties, it is necessary to calculate how the potential barrier $V_b$ changes in the GB upon applying gate voltage $V_g$. Top panel of Fig. 5 shows a TFT structure with a SGB located at the middle of the ZnO layer with Si$_3$N$_4$ gate insulator. In this case, we consider the values of $N_t$ in GB and $N_d$ for ZnO layer are $5.0 \times 10^{11}$ cm$^{-2}$ and $1.0 \times 10^{16}$ cm$^{-3}$, respectively. Figure 5(a) shows the $V_b - V_g$ curves at three locations in the GB along the ZnO thickness ($z = 0$, 50, and 100 nm). At $V_g = 0$, change in $V_b$ at three locations is only the effect of band bending due the work function difference between ZnO and Al gate electrodes.[28,35] As $V_g$ increases, the carriers are accumulated at ZnO-Si$_3$N$_4$ interface ($z = 0$), and $V_b$ reduces rapidly. Whereas, at other two locations ($z = 50$ and 100 nm), $V_b$ reduces slowly and becomes independent of $V_g$. This is due to the fact that the effective field developed by $V_g$ does not propagate deeply in the channel. Figure 5(b) shows the change in $V_b$ along the thickness of the ZnO film at GB for three gate voltages ($V_g = 0$, 5, and 10 V). At $V_g = 0$, $V_b$ remains high along the whole thickness, to make a conduction barrier for the carriers. At $V_g = 5$ and 10 V, $V_b$ reduces at ZnO-Si$_3$N$_4$ interface significantly to open up a channel for carrier conduction. These are the fundamental mechanisms of a field effect transistor consisted with a polycrystalline semiconductor. We also show a three-dimensional (3D) illustration of a calculated band bending profile in Fig. 6 to visualize the combined effect of Fig. 5(a) and (b). Three parts of Fig. 6 represent the conduction band profile of a SGB and its neighboring defect-free crystallite regions at the same instant for three gate bias conditions. The Fermi level for whole region of the ZnO film corresponds to the basal flat plane. The profile along the $x$ direction at ZnO-Si$_3$N$_4$ interface ($z = 0$) illustrates the real view of an actual TFT channel in presence of a single GB, while the profile along $z$ direction (ZnO thickness) illustrates bending profile from the ZnO-Si$_3$N$_4$ interface to a back channel of a TFT. The illustrations give graphical interpretation of a measure of an effective channel width along the active layer thickness. The region at which $V_b + V_n$ is comparable to $kT/q$ can be considered as an effective channel depth from ZnO-Si$_3$N$_4$ interface ($z = 0$). This simple SGB modeling shows a clear insight of trap activity in the GB for a TFT operation and redirects us to model the actual TFT consisting of multiple GBs.

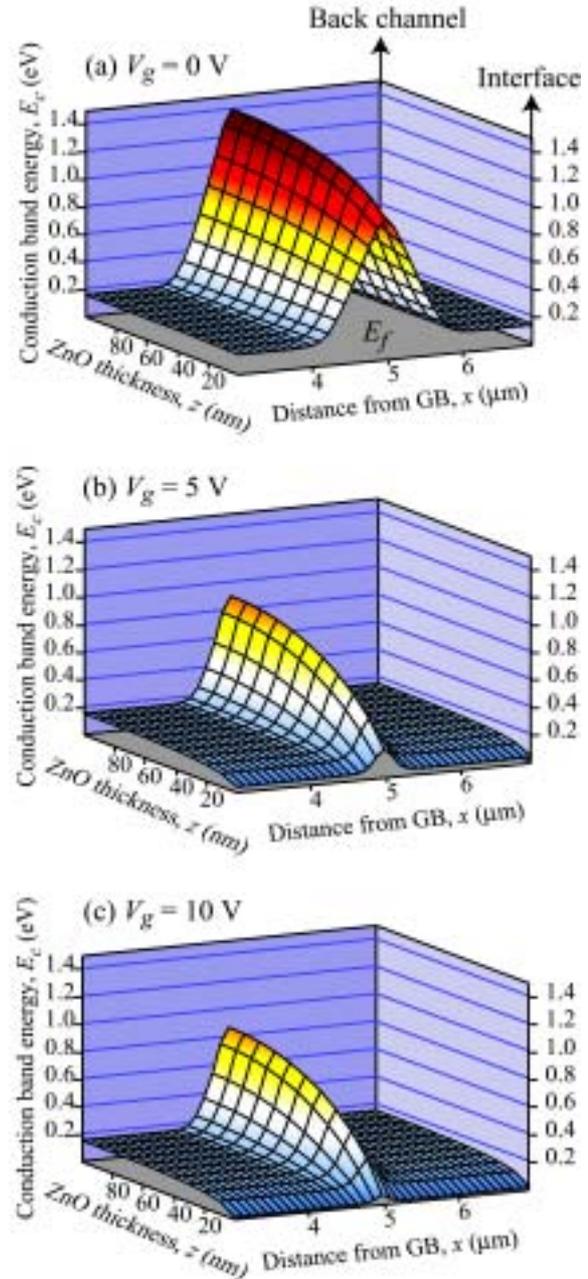

FIG. 6. 3D illustration of calculated band bending profile in and around single grain boundary (SGB) of ZnO thin film for three different gate bias conditions, (a) gate voltage $V_g = 0$ V, (b) $V_g = 5$ V, and (c) $V_g = 10$ V.

**B. Multiple grain boundary modeling**

To demonstrate the actual device modeling for the polycrystalline ZnO TFTs, we simulate the effect of multiple GBs. The model of multiple GBs is shown in Fig. 1 when many GBs (dotted lines) are considered together with SGB (solid line). We find that the behavior of GB barrier potential and band bending profile for multiple GBs are different from that of the SGB case. All GBs are placed in such a way that they are equally spaced i.e. the grain size $l_g$ over the entire ZnO film is a constant. Keeping the trap state density in all GBs same as in the SGB, developed symmetric barrier potentials at each GB are significantly overlapped one another, especially, when a condition of $l_g < 2\lambda$ is satisfied. Figure 7 shows a potential profile of a polycrystalline ZnO thin film for three different grain sizes. For all three cases, the values of $N_d$ and $N_t$ are taken $1.0 \times 10^{16}$ cm$^{-3}$ and $1.0 \times 10^{12}$ cm$^{-2}$, respectively. In Fig. 7(a), GBs are isolated from each other i.e. the depletion region associated with each GB are not overlapped and the minimum conduction band energy in the crystallite remains equal to $qV_n$, which is related with the free carrier concentration $n$ and effective density of states in the conduction band $N_c$, and represented as,

$$V_n = \frac{kT}{q}\ln\frac{N_c}{n} \quad (18)$$

when the condition $n_t (cm^{-2}) \leq N_d (cm^{-3}) l_g (cm)$ is satisfied. In this case, the carrier transport in the film is limited by the thermionic emission over the GB barrier potential. It shows the actual view of a microcrystalline thin film, usually observed in poly-Si thin film with a large grain. In Fig. 7(b), GBs are closer than that of Fig. 7(a) and the GB depletion regions are overlapped to lift the minimum conduction band edge $E_c$ up from the equilibrium Fermi level $E_f$ resulting in the activation energy $E_a$ higher than the $qV_n$. Consequently, the concentration of thermally activated carriers becomes smaller. Further reduction of grain size, as shown in Fig. 7(c), results in lifting up the $E_c$ further away from the $E_f$, and then the depletion region overlaps to spread over the entire film. On the other hand, the GB barrier height $V_b$ reduces, because $V_b$ is defined as $E_{cgb} - E_c$ (Fig. 2) i.e. the energy difference between the conduction band edge minima $E_c$ in crystallite and conduction band edge maxima $E_{cgb}$ in GB. This depicts a realistic potential profile for a nanocrystalline thin film and their properties by reducing $l_g$ or increasing the number of GBs. Hence, one may recognize the distinguishing features between a large grain microcrystalline poly-Si thin film and small grain nanocrystalline ZnO thin film for TFT action if these films are used as an active layer in the TFT structure. Therefore, it is important here to note that the grain size, donor concentration, and GB trap state densities are the important parameters for an *n*-type polycrystalline ZnO thin film to determine the properties of the ZnO thin film for TFT action self-consistently. We have already discussed the effect of GB potential upon changing the trap state density in the SGB case (see section III-A). Here we show how the properties of the thin film change for a fixed number of GBs from source to drain to get a transistor action by changing donor concentration $N_d$, which is analogous to the carrier modulation on applied gate bias in a TFT structure.

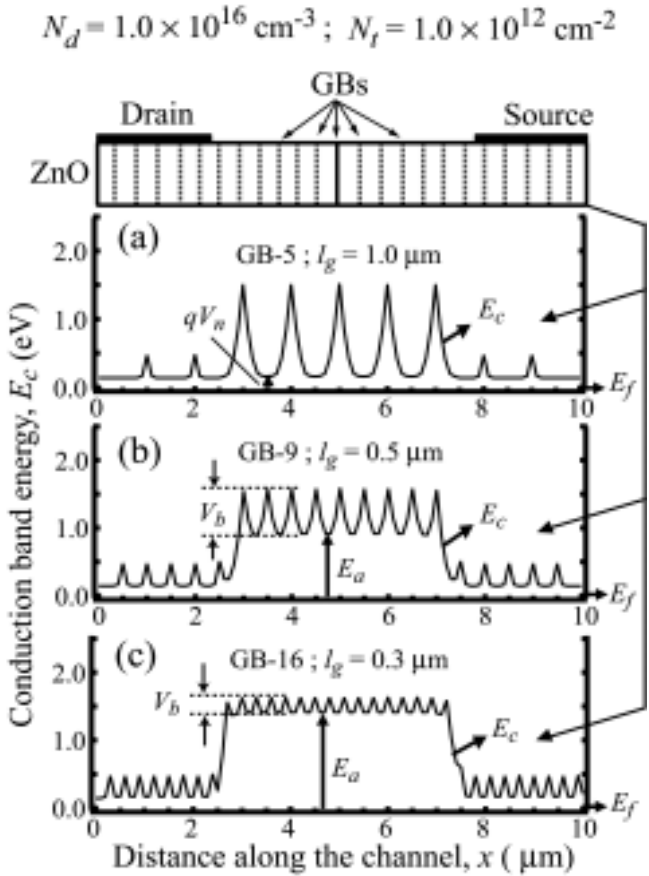

FIG. 7. Potential profile along the surface of ZnO thin film opposite to the source-drain electrodes for three different grain sizes (a) grain size $l_g = 1.0$ μm, corresponding effective number of GBs $n_g = 5$ (GB-5), (b) grain size $l_g = 0.5$ μm, $n_g = 9$ (GB-9), and (c) grain size $l_g = 0.3$ μm, $n_g = 9$ (GB-16).

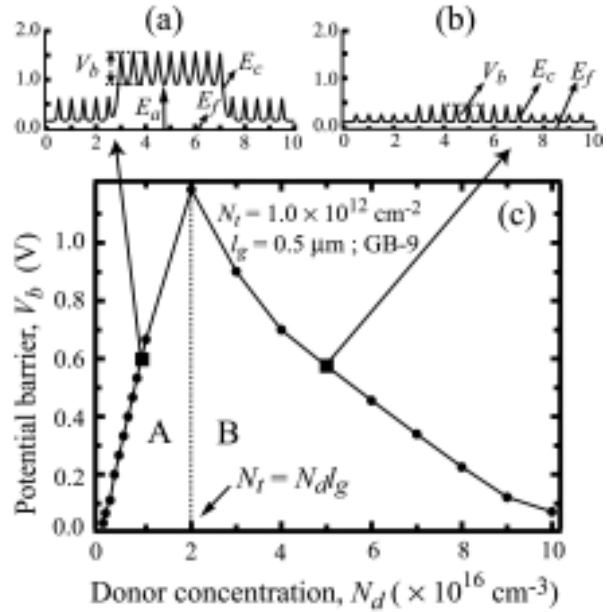

FIG. 8. Plot of GB barrier height $V_b$ against the donor density $N_d$, showing how $V_b$ reaches at its maximum value when $N_t = N_d l_g$ for grain size $l_g = 0.5$ μm (GB-9) and GBs potential profiles at two points in regions A and B of $V_b - N_d$ plot.

Figure 8(c) shows the donor concentration $N_d$ dependence of barrier potential $V_b$ for a fixed grain size $l_g$ and a total trap state density $N_t$ at GBs. The barrier height reaches a maximum at a critical donor density of $N_d = n_t / l_g$. The illustration can be easily divided into two parts (A and B), which are separated by a dotted line. The calculated potential profiles at two representative $N_d$ values [(a) $1.0 \times 10^{16}$ and (b) $5.0 \times 10^{16}$ cm$^{-3}$] are represented in Fig. 8(a) and (b), respectively. In region A, the depletion regions for each GB are overlapped and higher activation energy $E_a$ develops, whereas GBs become isolated in region B to make the grains partially depleted. The barrier potential for two regions can be represented by;[36]

$$V_b = \frac{qN_d l_g^2}{8\varepsilon_s} \quad (N_d l_g \leq n_t) \ (\infty \ N_d) \quad (19)$$

$$V_b = \frac{qn_t^2}{8\varepsilon_s N_d} \quad (N_d l_g \geq n_t) \text{ (inversely } \infty \ N_d) \quad (20)$$

Here, we evaluate $V_b$ as a function of $N_d$, that mimic the variation of $V_b$ as a function of $V_g$ in TFT structure because both parameters of $N_d$ and $V_g$ are correlated through the shift of Fermi level. Therefore, the Fig. 8(a) and (b) corresponding to region A and B represent the potential profiles in the channel of a TFT for off-state and on-state conditions, respectively. The analytical solutions of on-state conductivity or mobility can be obtained for a TFT in presence of GBs using equation (20) by replacing the donor density $N_d$ with gate bias assisted accumulation of free carrier density $n$.

**C. Device characteristics**

In light of above modeling and discussions, we have simulated the TFT characteristics taking ZnO as an active material for single crystal (no GB) and poly-crystal (with GBs) cases for which schematic device structures are given in Fig. 1. Figure 9(a) shows the transfer characteristics of a single crystal ZnO TFT with a constant $V_{ds} = 10$ V. A sharp rise of sub-threshold slope and above threshold linearity of $\sqrt{I_d} - V_g$ characteristics indicates the perfect saturation behavior of ZnO channel. The field effect mobility $\mu_{FE}$ and the threshold voltage $V_{th}$ are calculated using conventional method *i.e.* from the slope and the *x*-axis intercept of $\sqrt{I_d} - V_g$ curve, respectively. The calculated field effect mobility $\mu_{FE}$ is 142 cm$^2$/Vs, which is very close to our considered materials constant (150 cm$^2$/Vs) listed in Table I. This implies that the device simulator can perfectly characterize the TFT properties suggesting that our model for device characterization is acceptable. Then the GBs are inserted in the ZnO layer as represented in Fig. 1 to make it poly crystalline and then the transfer characteristics are obtained for 9 and 16 GBs within the channel length of 5 μm as shown in Fig. 9(b). Here, the sub-threshold slope shows a considerable tailing, and $\sqrt{I_d} - V_g$ curve deviate from linear relationship above threshold, besides, the threshold voltage shifts to the positive direction of gate voltages. All these effects are more enhanced for more number of GBs. The $\mu_{FE}$ calculated from maximal slope for GB-9 and GB-16 are 91 cm$^2$/Vs and 63 cm$^2$/Vs, respectively. In a similar manner, we can extract above threshold mobility for many GBs of nanocrystalline ZnO based TFTs.

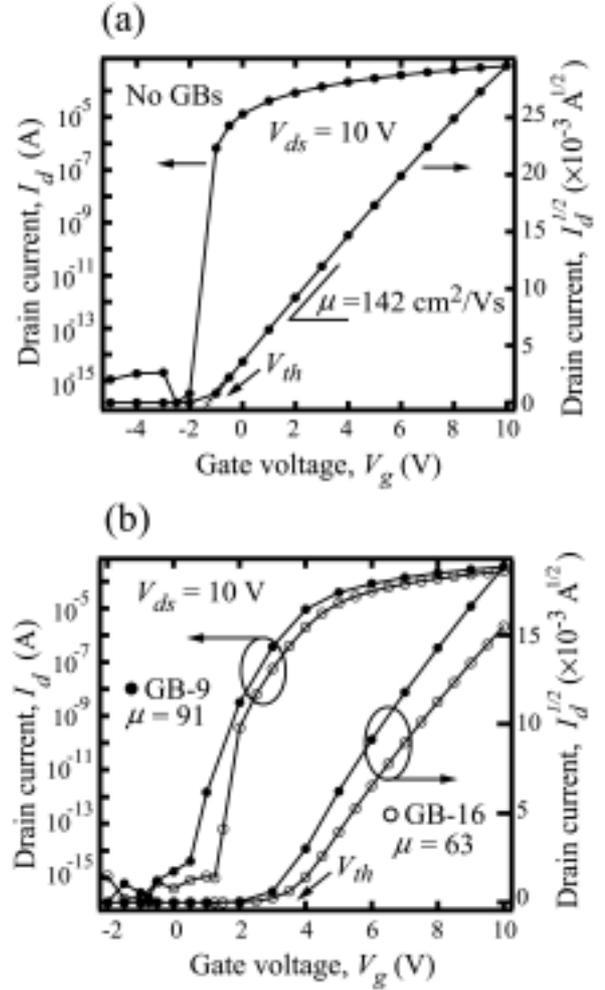

FIG. 9. Transfer characteristics of a modeled ZnO-TFT (a) without any GBs (single crystal ZnO TFT) (b) with effective number of GBs (GB-9 and GB-16) within the channel length of ZnO TFT.

**D. Extraction of the GB trap state density**

As the most prominent part of this paper, here, we propose an analytical way to extract the GB trap state density from experimentally obtained TFT characteristics. As far as we are concerned with the above threshold mobility, we need to observe the actual potential profiles in a channel when gate voltage $V_g$ is applied above the threshold voltage $V_{th}$. Figure 10 elucidates the potential profile at ZnO-Si$_3$N$_4$ interface, where accumulation channel is formed for three different grain sizes (1.0, 0.5, and 0.3 μm) corresponding to GB-5, GB-9 and GB-16, respectively. In all three cases, mobile carriers from source to drain will only face the small potential peaks comparable to $kT/q$ along its direction of propagation. If we apply low drain-source voltage of $V_{ds} = 0.1$ V which is far lower than $V_g$, almost all GBs remain effective within the channel length $L$, where we call linear region. For a relatively high $V_{ds}$ compared with $V_g$ (for instance, $V_{ds} = 10$ V and $V_g = 5$ V), a few GBs near drain end are diminished due to pinch-off, but the remaining GBs cause to limit the mobility, where we call saturation region. The simulator can analytically derive the relationship between mobility and number of GBs assuming no voltage drop between each GB (in the crystallite). The assumption here is quite reasonable, because the resistance in crystallite region at above threshold is much lower than that in the GB region. To establish the relationship, the following equations are used[37,38].

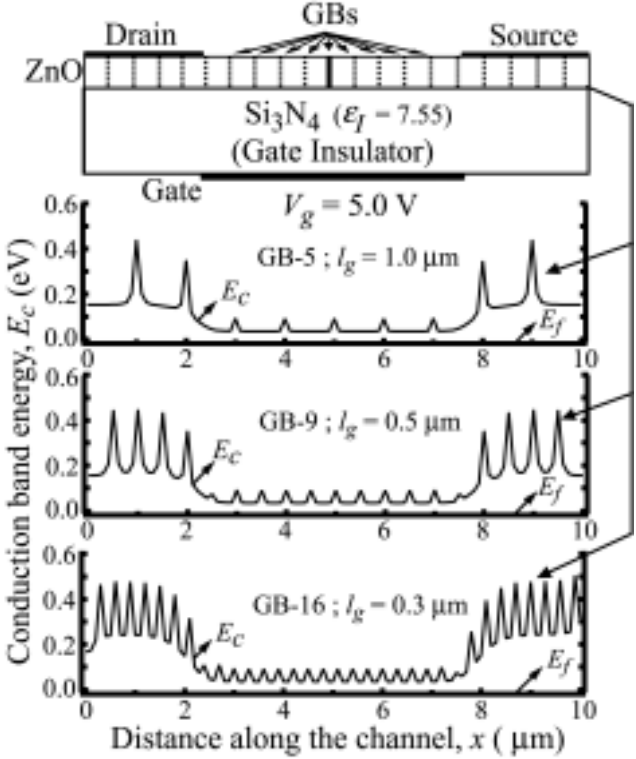

FIG. 10. Potential profile along the ZnO-Si$_3$N$_4$ interface, when gate bias ($V_g = 5.0$ V) is above the threshold for three different grain sizes $l_g = 1.0$, 0.5, and 0.3 μm corresponding to GB-5, GB-9, and GB-16, respectively within the channel length $L$.

Current density through the grain boundary barrier is expressed as

$$J_{gb} = qv_n n \exp(-qV_b/kT)[\exp(q\Delta V_b/kT) - 1] \quad (21)$$

where $v_n = AT^2/qN_c$ is the thermal velocity of electron and $\Delta V_b$ is the voltage drop across each GB. Equation (21) replicates the equations (12) and (13), thus it includes both the drift-diffusion and thermionic emission effects of the transport. If the drain bias is high enough to satisfy $q\Delta V_b \gg kT$, equation (21) becomes,

$$J_{gb} = qv_n n \exp(-qV_b/kT) \exp(q\Delta V_b/kT) \quad (22)$$

If there are $n_g$ grain boundaries within the channel length $L$, then $\Delta V_b$ equals to $V_{ds}/n_g$. For a uniform electric field in the channel, the macroscopic (source to drain) conductivity is expressed by,

$$\sigma = J_{gb}L/V_{ds} = \frac{qv_n nL}{V_{ds}} \exp(-qV_b/kT) \exp(qV_{ds}/n_g kT) \quad (23)$$

Equation (23) can be simplified as,

$$\sigma = qn\mu^* \quad (24)$$

Where $\mu^*$ is a reduced effective mobility, which is limited by the number of GBs $n_g$ in the channel and the GB potential barrier $V_b$.

$$\mu^* = \frac{v_n L}{V_{ds}} \exp(qV_{ds}/n_g kT) \exp(-qV_b/kT) \quad (25)$$

Note that the term $\exp(-qV_b/kT)$, associated with each GB, is gate bias $V_g$ dependent. At above threshold $V_g = 5$ V, $V_b$ reduces so that it becomes comparable to $kT/q$ for all GBs in the channel (see Fig. 10) and then the contribution of this exponential term to control mobility becomes small for $V_g$ ranges 5 to 10 V, where we conventionally calculate the field effect mobility $\mu_{FE}$. Another exponential term $\exp(qV_{ds}/n_g kT)$ remains effective and controls $\mu^*$ depending on the number of GBs $n_g$ in the channel, because it integrates all the small contributions of the term $\exp(-qV_b/kT)$ associated with it as mentioned above. Thus, $\mu^*$ in equation (25) represents the field effect mobility $\mu_{FE}$ at the on-state condition of a TFT. Therefore, it can be stated that the effective number of GBs inside the channel length limits the field effect carrier mobility of the TFTs at its on-state condition. The analytical derivation of equation (25) implies that $\mu_{FE}$ exponentially decreases with increasing the number of GBs $n_g$.

Finally, we have calculated the field effect mobility from the transfer characteristics of our modeled TFT for the effective number of GBs in the channel considering three different total trap state densities $N_t$ in the GBs as elucidated in Fig. 11. Five points are actual simulated data points for three $N_t$ values. Extensions of these data points are the exponential extrapolation based on equation (25). It is clearly seen that those five points are sufficient for exponential curve fitting. We extend these three curves up to the experimental data range, which are usually extracted from our several ZnO-based TFTs.[8,39] Low temperature (150°C-500°C) deposited nanocrystalline ZnO films exhibit number of GBs ranging from 50 to 100 within the channel length $L = 5$ μm and the field effect mobilities ranging from 1 to 10 cm$^2$/Vs for a wide range of gate biases. The crossing arrows in Fig. 11 shows the experimental data ranges. If we compare our experimental data with this view graph, it is possible to make a rough estimation of the GBs areal total trap state density, $N_t$ which is of the order of $10^{12}$ cm$^{-2}$ for our ZnO-based TFTs. This value is quite realistic, about 1% of the total number of surface atoms of a ZnO thin film, which is of the order of $10^{14}$ cm$^{-2}$. Therefore, one can treat this view graph as a general picture for a rough estimation of GBs areal trap states densities of any ZnO-based TFTs as soon as one knows their co-ordinate (mobility as ordinate and effective number of GBs as abscissa) point. A recent report of Carcia et al.[7] on rf magnetron sputtered ZnO TFT, estimated GBs trap densities ranging from $1.5 \times 10^{12}$ cm$^{-2}$ to $3.0 \times 10^{12}$ cm$^{-2}$ and their estimated field effect mobility is 1.2 cm$^2$/Vs for one of their good TFT. This result is also comparable to our simulation result. Hence, we emphasize that our simulation method for ZnO-based TFT is exclusively consistent with the experiment.

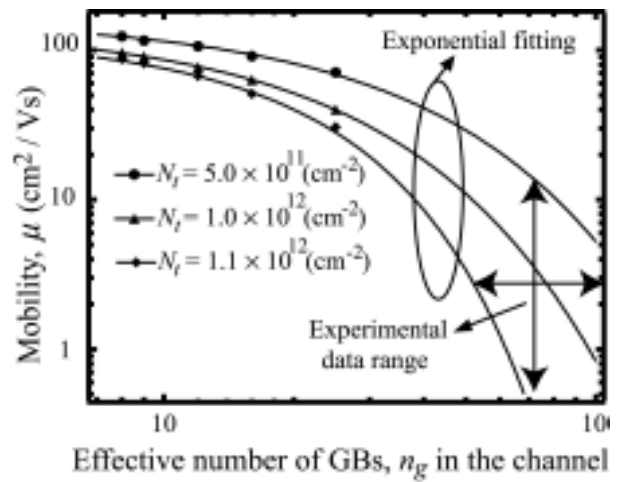

FIG. 11. Calculated mobility $\mu_{FE}$ vs. effective number of GBs $n_g$ plot for three different total areal trap state density $N_t$ values and the exponential extrapolations (equation (25)) up to the experimental data ranges.

## IV. CONCLUSION

Device simulation of ZnO TFTs was performed using the device simulator. An actual polycrystalline ZnO TFT modeling was commenced with considering a SGB positioned at the middle of the ZnO layer in a TFT structure. A double Shottky barrier formation at the GB adequately described the nature of potential barrier and the spreading of depletion region from the GB to the crystallite for a wide range of trap state densities localized at the GB. The SGB model was also suitably described the change of potential profile in and around the GB with applied gate electric field, exhibits how effectively the channel is formed across the GB. The multiple GB model afterward facilitated us to observe the real microscopic view of a developed potential profile in the nanocrystalline ZnO active layer. We competently explained the action of a nanocrystalline ZnO TFT using our calculated potential profiles. Apparently, our assumption on the trap distribution in the band gap is too simple and crude. Nonetheless, it is good enough to visualize the potential profile in the channel upon gate bias application. The most promising option of this simulation is to calculate the approximate range of trap state densities localized in GBs comparing with experimentally obtained I-V characteristics of any polycrystalline ZnO TFTs, without executing spectroscopic measurement.

Next step would be to analyze the trap distribution through measurement of temperature dependence for the TFT characteristics and compare them with simulation results. That gives more detailed insight to the trap distribution. Then, one can test some of the chemical modification of the GBs and extract the effects on how the trap distribution is modified.

## ACKNOWLEDGMENTS

This work was supported by the MEXT Grant of Creative Scientific Research #14GS0204, the Nissan Science Foundation, and Nippon Sheet Glass Foundation for Materials Science and Engineering.